\documentclass[onecolumn,amsmath,amssymb,superscriptaddress,nofootinbib,11pt,a4paper]{revtex4}
\usepackage{titlesec}
\titleformat{\section}
  {\normalfont\Large\bfseries}
  {\thesection}
  {1em}
  {}
\titleformat{\subsection}
  {\normalfont\large\bfseries}
  {\thesubsection}
  {1em}
  {}
\pdfoutput=1
\usepackage[T1]{fontenc}
\usepackage{CJK}
\usepackage{xcolor}
\usepackage{amsfonts}
\usepackage{epstopdf}
\usepackage{wrapfig}
\usepackage{subcaption}
\usepackage{graphicx}
\usepackage{dcolumn}
\usepackage{bm}
\usepackage{float}
\usepackage[utf8]{inputenc}
\usepackage{booktabs}
\usepackage{orcidlink}
\usepackage{multirow}

\begin{document}

\title{Extracting Energy from a Traversable Wormhole via Magnetic Reconnection: Circular Orbits and the Plunging Region}

\author{Ke Wang}
\email{kkwwang2025@163.com}
\affiliation{College of Physics and Optoelectronic Engineering, Chongqing Normal University, \\Chongqing 401331, China}
\affiliation{School of Material Science and Engineering, Chongqing Jiaotong University, \\Chongqing 400074, China}

\author{Wei Xu}
\email{xwcqnu@163.com}
\affiliation{College of Physics and Optoelectronic Engineering, Chongqing Normal University, \\Chongqing 401331, China}

\author{Xiao-Xiong Zeng}
\email{xxzengphysics@163.com (Corresponding author)}
\affiliation{College of Physics and Optoelectronic Engineering, Chongqing Normal University, \\Chongqing 401331, China}

\begin{abstract}
Using the Comisso-Asenjo magnetic reconnection mechanism, we study energy extraction from the Teo traversable rotating wormhole at different orbital locations. This wormhole  has no event horizon and cannot reduce to a Kerr black hole in any parameter limit. We compute circular orbits on the equatorial plane, the prograde photon sphere, and the innermost stable circular orbit (ISCO). We find that below a certain spin, no circular orbit exists inside the ergosphere. Both the efficiency and power calculated from magnetic reconnection increase monotonically with angular momentum. Compared with a Kerr black hole under the same plasma parameters, the Teo wormhole has higher energy extraction power and efficiency. This is completely different from the Simpson-Visser rotating wormhole, which can reduce to a Kerr black hole, where the large  spin makes efficiency and power decrease. We also extend magnetic reconnection to the plunging region. We find that when the spin is not particularly high, the energy extraction power and efficiency in the plunging region are higher than those in the circular orbit region, while for very high spin the situation is reversed. This conclusion also differs from that obtained from a Kerr black hole, where the power and efficiency in the plunging region are always higher than those in the circular orbit region.
\end{abstract}
\maketitle

\newpage
\section{Introduction}
A wormhole is a special solution of the Einstein field equations. It describes a spacetime channel connecting two asymptotically flat regions, or two distant regions of the same universe. Since the work of Einstein and Rosen \cite{4}, wormholes were long regarded as purely theoretical objects, until Morris and Thorne \cite{1} systematically proposed the concept of a traversable wormhole. People then realized that as long as matter violating the null energy condition is allowed, the field equations permit a spacetime structure that has no event horizon, no curvature singularity, and can be traversed in both directions. Compared with black holes, traversable wormholes can be highly similar in external observables, such as shadows \cite{23}, photon rings \cite{35}, quasinormal modes \cite{24}, and accretion disks \cite{36}. Therefore wormholes are also regarded as an important class of black hole mimickers \cite{22}.

How to efficiently extract energy from the ergosphere of a rotating black hole has long been a topic explored by physicists. Penrose \cite{10} first pointed out that when a particle splits into two branches inside the ergosphere, and one branch carries negative energy at infinity and falls into the black hole, the other branch can escape to infinity carrying more energy than the original particle. This is the famous Penrose process. However, this mechanism has some drawbacks, such as the requirement of a very high relative velocity between the separated particles \cite{25}. This gave rise to other energy extraction mechanisms \cite{26,27,28,29,30,31,32,33,34,6}. Among these mechanisms, the Comisso-Asenjo mechanism \cite{6} is representative and easy to calculate quantitatively. Based on the magnetic reconnection mechanism, Comisso and Asenjo found in Kerr black holes that the energy extraction power is higher than that of the traditional Blandford-Znajek mechanism \cite{29}. For this reason, the magnetic reconnection mechanism has been rapidly extended to various black hole backgrounds in recent years \cite{37,38,39,40,41,42,43,44,45,46,47,48,50,51,60,61,63,64,65,66,67,71}. In particular, references \cite{52,53} first extended magnetic reconnection to the plunging region. The results showed that in Kerr black holes the energy extraction efficiency in the plunging region is higher than that in the circular orbit region. This motivated studies of magnetic reconnection in the plunging region for different black hole backgrounds \cite{49,54,55,56,57,58,59,62,68,69,70}. In addition, people have used the Comisso-Asenjo mechanism to combine magnetic reconnection with hotspot imaging, attempting to identify energy extraction signals in astronomical observation data \cite{13,14,15,16,17}. At the same time, spacetime curvature itself can significantly modulate the properties of magnetic reconnection \cite{18,19,20}, and self-consistent numerical simulations have also confirmed that the Penrose process driven by magnetic reconnection is indeed effective under realistic astrophysical conditions \cite{21}.

At present, the Comisso-Asenjo mechanism has been widely extended to black holes, while studies of the Comisso-Asenjo mechanism in wormholes are very few, except for reference \cite{3}. Moreover, that work did not study magnetic reconnection energy extraction in the plunging region. That paper studied a rotating Simpson-Visser type wormhole \cite{72,73}. This spacetime is only a minimal regularization of the Kerr spacetime. In the limit $\ell\to0$ it exactly reduces to a Kerr black hole, and in the limit $a\to0$ it returns to a static wormhole. The horizon, ergosphere, photon sphere, and ISCO are all obtained from the corresponding Kerr formulas by the replacement $r\to\sqrt{r^{2}+\ell^{2}}$. For rapid rotation there even exist inner and outer ergosurfaces. Based on this, reference \cite{3} obtained two important conclusions. The wormhole spin in the allowed region for energy extraction is restricted to a certain range, and energy cannot be extracted if the spin is too high or too low. Both efficiency and power decrease as spin increases, and the efficiency also exhibits a nonmonotonic peak with respect to the regularization parameter $\ell$. These conclusions clearly bear the deep imprint of the Kerr algebraic structure.

This paper studies a class of rotating wormholes that has no relation to the Kerr algebraic structure, namely the Teo traversable rotating wormhole \cite{2}. In 1998, Teo extended the Morris-Thorne wormhole to the stationary axisymmetric case and gave the regular form of the rotating traversable wormhole metric. Currently, many works in the literature have studied Teo wormholes \cite{75,76,77}. This model has a series of properties which are completely different from Kerr type wormholes, such as having no Kerr limit and having no upper bound on spin. These structural differences immediately raise a natural question. In such a wormhole spacetime, will energy extraction via magnetic reconnection behave like a Kerr black hole, or like the case in reference \cite{3} where the large spin is instead unfavorable? On the other hand, the situation of magnetic reconnection in the plunging region of a wormhole is currently unclear. Does the magnetic reconnection process in the plunging region of a wormhole show the same behavior as in a Kerr black hole? Are the power and efficiency in the plunging region still higher than those in the circular orbit region? We will explore these questions in this paper.

The structure of this paper is as follows. Section 2 introduces the Teo wormhole. Section 3 gives circular orbits, the photon sphere, and the ISCO on the equatorial plane. Section 4 establishes the zero angular momentum observer (ZAMO) framework and analyzes the magnetic reconnection energy extraction process. Section 5 calculates the power and efficiency of energy extraction and compares them with the Kerr black hole and the rotating wormhole in reference \cite{3}. Section 6 extends the reconnection region to the plunging region. The final section gives conclusions.

\section{Brief Introduction to the Teo Wormhole}

Morris and Thorne first proposed the concept of a traversable wormhole in 1988 \cite{1} with  a reverse approach. They first listed the geometric conditions that a traversable wormhole should satisfy, and then used the Einstein field equations to infer the matter distribution needed to support the wormhole. Teo extended this framework to the stationary axisymmetric case in 1998 and gave the regular form of the traversable rotating wormhole metric \cite{2}
\begin{equation}\label{Teometric}
ds^{2}=-N^{2}dt^{2}+\left(1-\frac{b}{r}\right)^{-1}dr^{2}+r^{2}K^{2}\left[d\theta^{2}+\sin^{2}\theta\left(d\varphi-\omega dt\right)^{2}\right],
\end{equation}
where the four gravitational potentials $N$, $b$, $K$, and $\omega$ are functions of $r$ and $\theta$, and satisfy regularity conditions on the symmetry axis $\theta=0,\pi$, that is, their derivatives with respect to $\theta$ vanish on the axis. This metric describes two identical asymptotically flat regions smoothly joined at the throat $r=b>0$. For this geometry to indeed describe a traversable wormhole, the functions also need to satisfy the following conditions. (1) $N$ is finite and nonzero everywhere, so there is no event horizon and no curvature singularity. (2) The shape function satisfies $b\leq r$, and at the throat $b_{\theta}=0$ and the flare-out condition $b_{r}<1$ is satisfied. (3) $K$ gives the proper radial distance through $R\equiv rK$. (4) $\omega$ is the angular velocity of frame dragging. This paper adopts the simplest model in this regular form that also carries Arnowitt-Deser-Misner (ADM) mass
\begin{equation}\label{ansatz}
N^{2}=1-\frac{2M}{r},\quad b=r_{0},\quad K=1,\quad \omega=\frac{2J}{r^{3}}=\frac{2aM}{r^{3}},
\end{equation}
where $M$ is the ADM mass, $J$ is the total angular momentum, $a\equiv J/M$ is the specific angular momentum, the dimensionless spin is $a/M=J/M^{2}$, and $r_{0}$ is the throat radius. Traversability requires
\begin{equation}\label{r0bound}
r_{0}>2M .
\end{equation}

The above choice is not arbitrary. The reasons are as follows. (i) $N^{2}=1-\frac{2M}{r}$. First, it guarantees the asymptotic expansion $g_{tt}=-1+2M/r+O(r^{-4})$, so that $M$ is exactly the ADM mass. Second, the zero of $N$ is at $r=2M$, so as long as we take (\ref{r0bound}), $N$ is finite and nonzero everywhere in the region $r\geq r_{0}$ where the wormhole is located. There is neither an event horizon nor a curvature singularity. This is exactly Teo's requirement for traversability. Finally, in the nonrotating limit $a\to0$, this choice makes the spatial slice reduce to the well known Einstein-Rosen bridge, that is, a Schwarzschild slice wormhole \cite{4,5}. It can be regarded as a traversable regularization of that bridge when $r_{0}>2M$, which is convenient for comparison with the static case. (ii) $b=r_{0}$. This is the simplest shape function. Since $b_{\theta}\equiv0$, the throat must be located at constant $r=r_{0}$, and the $(r-b)^{-2}$ type curvature singularity in equation (9) of Teo \cite{2} automatically disappears. At the same time, the flare-out condition
\begin{equation}
\frac{b-rb_{r}}{2b^{2}}\bigg|_{r=r_{0}}=\frac{1}{2r_{0}}>0
\end{equation}
always holds. More importantly, because $b$ is independent of $\theta$, the intrinsic radial coordinate $l$ defined in equation (15) of reference \cite{2} is no longer only a first order approximation near the throat, but is globally valid and strictly covers the entire spacetime. (iii) $K=1$. In this case $g_{\theta\theta}=r^{2}$ and $g_{\varphi\varphi}=r^{2}\sin^{2}\theta$, that is, $r$ is the area radius, a constant $r$ surface is a spherical surface of radius $r$, and the area of the throat is $4\pi r_{0}^{2}$. This also means that the angular part is exactly the same as in Schwarzschild spacetime, and all rotation is concentrated in $\omega$ in $g_{t\varphi}$. Therefore (\ref{ansatz}) can be regarded as a minimal rotation model obtained by superposing pure frame dragging on the Einstein-Rosen bridge \cite{4,5}. (iv) $\omega=2J/r^{3}$. Teo's asymptotic condition requires $\omega=2J/r^{3}+O(r^{-4})$ \cite{2}, where the coefficient $2J$ of the leading term guarantees that $J$ is exactly the total angular momentum of the spacetime.

From (\ref{ansatz}), $b=r_{0}$ is constant, so the throat is located at $r=r_{0}$. The throat is a spherical surface. Its intrinsic metric is $r_{0}^{2}(d\theta^{2}+\sin^{2}\theta d\varphi^{2})$, its area is $4\pi r_{0}^{2}$, and its equatorial circumference is $2\pi r_{0}$. The ergosphere is defined by $g_{tt}>0$, and its boundary is given by $g_{tt}=0$. That is,
\begin{equation}\label{ergeq}
g_{tt}=-\left(1-\frac{2M}{r}\right)+\frac{4J^{2}\sin^{2}\theta}{r^{4}}=0
\quad\Longleftrightarrow\quad
r_{\rm erg}^{3}\left(r_{\rm erg}-2M\right)=4J^{2}\sin^{2}\theta=4a^{2}M^{2}\sin^{2}\theta .
\end{equation}
Since
\begin{equation}
\frac{d g_{tt}}{dr}=-\frac{2M}{r^{2}}-\frac{16J^{2}\sin^{2}\theta}{r^{5}}<0
\end{equation}
holds for all $r>0$, $g_{tt}$ decreases monotonically with $r$ and $g_{tt}(\infty)=-1<0$. Therefore this model has at most one ergosurface at each $\theta$, and there is no inner ergosurface like that in Kerr spacetime. When $J\to0$, the ergosphere disappears. This is consistent with the fact that a static wormhole contains no ergosphere. In addition, the spin $a/M$ has no upper bound. This is completely different from the Kerr black hole restriction $|a|\leq M$, and it also echoes the conclusion in reference \cite{3} that the rotating wormhole allows $a>M$. For energy extraction, one needs to satisfy
\begin{equation}\label{ergoregion}
r_{0}\leq r<r_{\rm erg}.
\end{equation}
\begin{figure}[!h]
  \centering
  \begin{subfigure}{0.45\textwidth}
    \centering
    \includegraphics[width=\linewidth]{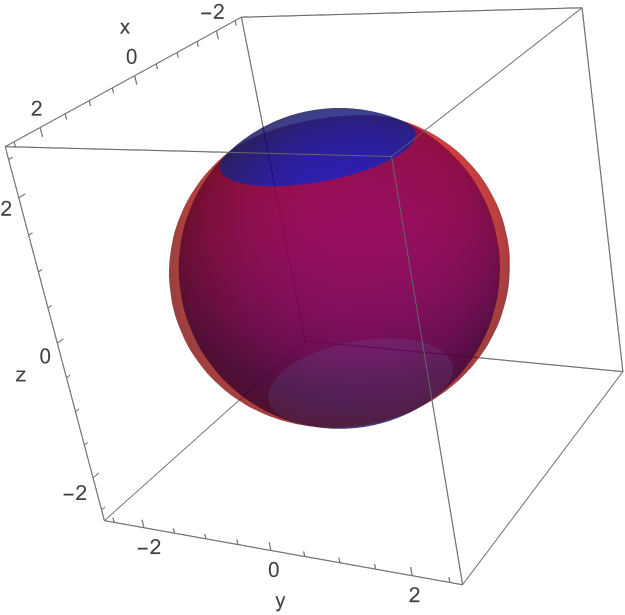}
    \caption{$a/M=0.8$}
  \end{subfigure}
  \begin{subfigure}{0.45\textwidth}
    \centering
    \includegraphics[width=\linewidth]{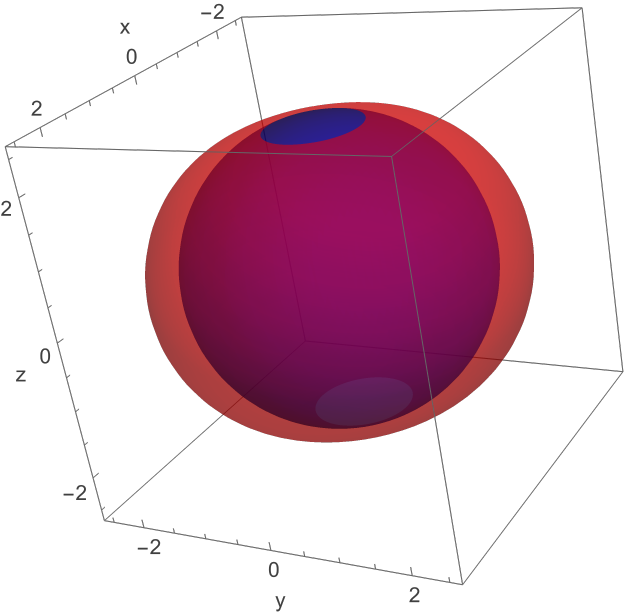}
    \caption{$a/M=1.5$}
  \end{subfigure}
\caption{The throat and ergosphere of the wormhole in $x-y-z$ three-dimensional space, with $r_0/M=2.1$. Blue is the throat, orange red is the ergosphere boundary.}
\label{fig:1}
\end{figure}
We plot the throat and ergosphere of the wormhole in figure \ref{fig:1}. It can be seen that there is no ergosphere in the polar regions of the wormhole. In addition, we give numerical values of the equatorial ergosphere boundary for several typical spins in table \ref{tab:erg}. The data show that the equatorial ergosphere boundary increases with spin.
\begin{table}[htbp]
\caption{Equatorial ergosphere radius $r_{\rm erg}^{\rm eq}$ for different spins, obtained from $r_{\rm erg}^{3}(r_{\rm erg}-2M)=4J^{2}$.}
\label{tab:erg}
\begin{ruledtabular}
\begin{tabular}{ccccccccc}
$a/M$ & 0.5 & 1.0 & 1.5 & 2.0 & 2.5 & 3.0 & 4.0 & 5.0 \\
\hline
$r_{\rm erg}^{\rm eq}/M$ & 2.107 & 2.320 & 2.546 & 2.761 & 2.962 & 3.151 & 3.497 & 3.809 \\
\end{tabular}
\end{ruledtabular}
\end{table}

It should be pointed out that the rotating wormhole studied in reference \cite{3} reduces to a Kerr black hole when the wormhole parameter approaches zero. However, the Teo wormhole used in this paper cannot reduce to a Kerr black hole. This is because the angular part of the metric $g_{\theta\theta}=r^{2}$ and $g_{\varphi\varphi}=r^{2}\sin^{2}\theta$ is exactly the same as in Schwarzschild spacetime and contains no Kerr $a^{2}$ correction.

\section{Circular Orbits and Characteristic Radii on the Equatorial Plane of the Teo Wormhole}

In this section, we calculate circular orbits, the photon sphere, and the innermost stable circular orbit on the equatorial plane. This is done as in \cite{6}, placing the magnetic reconnection process on the equatorial plane. Note that if the current sheet is located inside the ergosphere, only prograde orbits exist there. Therefore in the following we discuss only the prograde branch.

Let $p_{t}=-E$ and $p_{\varphi}=L$. From $g^{\mu\nu}p_{\mu}p_{\nu}=-\mu^{2}$, where $\mu^{2}=0$ and $1$ correspond to null and timelike geodesics respectively, we obtain
\begin{equation}\label{HJ}
-\frac{\left(E-\omega L\right)^{2}}{N^{2}}+\frac{1}{r^{2}}\left(p_{\theta}^{2}+\frac{L^{2}}{\sin^{2}\theta}\right)+\left(1-\frac{r_{0}}{r}\right)p_{r}^{2}=-\mu^{2} .
\end{equation}
Define
\begin{equation}\label{Carter}
\mathcal{K}\equiv p_{\theta}^{2}+\frac{L^{2}}{\sin^{2}\theta}\geq0 ,
\end{equation}
which plays the same role as the Carter constant and is a conserved quantity. On the equatorial plane $\theta=\pi/2$ and $p_{\theta}=0$, so $\mathcal{K}=L^{2}$. Therefore the radial equation and angular equation separate completely
\begin{equation}\label{radialeq}
\dot{r}^{2}=\left(1-\frac{r_{0}}{r}\right)\left[\frac{\left(E-\omega L\right)^{2}}{N^{2}}-\mu^{2}-\frac{\mathcal{K}}{r^{2}}\right] .
\end{equation}
At $r>r_{0}$, circular orbits are given by the conditions that the function in the bracket and its first derivative both vanish. The angular velocity of an equatorial circular orbit $\Omega=d\varphi/dt$ is determined by
\begin{equation}\label{Omcond}
g_{\varphi\varphi}'\Omega^{2}+2g_{t\varphi}'\Omega+g_{tt}'=0 ,
\end{equation}
where a prime denotes a derivative with respect to $r$. Substituting $N^{2}=1-2M/r$ and $\omega=2J/r^{3}$, we obtain
\begin{equation}\label{OmegaK}
\Omega_{K}^{\pm}=\frac{-J\pm\sqrt{Mr^{3}+9J^{2}}}{r^{3}},
\end{equation}
where the upper sign is prograde and the lower sign is retrograde. In this paper we take only the upper sign. For convenience of writing, introduce
\begin{equation}\label{defPQ}
\mathcal{P}\equiv\sqrt{Mr^{3}+9J^{2}},\qquad
\mathcal{Q}\equiv\mathcal{P}-3J>0 ,
\end{equation}
then $\Omega_{K}^{+}=\left(\mathcal{P}-J\right)/r^{3}$, and
\begin{equation}\label{Omminusomega}
\Omega_{K}^{+}-\omega=\frac{\mathcal{Q}}{r^{3}} .
\end{equation}
From $g_{\mu\nu}\dot{x}^{\mu}\dot{x}^{\nu}=-1$ and the identity $g_{tt}+2\Omega g_{t\varphi}+\Omega^{2}g_{\varphi\varphi}=-N^{2}+r^{2}\left(\Omega-\omega\right)^{2}$, we get $\dot{t}=r^{2}/\sqrt{\mathcal{D}}$, where
\begin{equation}\label{Dpm}
\mathcal{D}\equiv r^{4}-3Mr^{3}-18J^{2}+6J\mathcal{P} .
\end{equation}
The condition for the existence of a prograde circular orbit on the equatorial plane is
\begin{equation}\label{existr}
\mathcal{D}>0\quad\Longleftrightarrow\quad r>r_{\rm ph} ,
\end{equation}
where $r_{\rm ph}$ is given by (\ref{rphradical}) below. In other words, timelike circular orbits on the equatorial plane exist only outside the prograde photon sphere. At $r\leq r_{\rm ph}$, the $t$ component of the four-velocity diverges, and there is no Keplerian plasma. From $E=-\left(g_{tt}+\Omega g_{t\varphi}\right)\dot{t}$ and $L=\left(g_{t\varphi}+\Omega g_{\varphi\varphi}\right)\dot{t}$, we obtain the specific energy and specific angular momentum of a particle on a circular orbit
\begin{equation}\label{EL}
L=\frac{r\,\mathcal{Q}}{\sqrt{\mathcal{D}}},\qquad
E=\frac{r^{4}-2Mr^{3}+2J\mathcal{Q}}{r^{2}\sqrt{\mathcal{D}}} .
\end{equation}
When $J=0$, $\mathcal{P}=\sqrt{Mr^{3}}$ and $\mathcal{D}=r^{3}(r-3M)$, and the above expressions return to the Schwarzschild results $E=(r-2M)/\sqrt{r(r-3M)}$ and $L=\sqrt{Mr^{2}/(r-3M)}$.

For null circular orbits, $\mu^{2}=0$ and on the equatorial plane $\mathcal{K}=L^{2}$, $ds^{2}=0$ directly gives $\Omega=\omega+N/r$, the prograde branch. Combining this with (\ref{OmegaK}), the photon sphere condition can be written in a compact form containing only metric functions
\begin{equation}\label{phcond}
\frac{N}{r}-N'=r\,\omega' .
\end{equation}
Substituting $N=\sqrt{1-2M/r}$ and $\omega=2J/r^{3}$, and using
\begin{equation}
\frac{N}{r}-N'=\frac{r-3M}{Nr^{2}},\qquad r\,\omega'=-\frac{6J}{r^{3}} ,
\end{equation}
equation (\ref{phcond}) becomes
\begin{equation}\label{rphradical}
r_{\rm ph}\left(r_{\rm ph}-3M\right)=-\,6J\sqrt{1-\frac{2M}{r_{\rm ph}}} ,
\end{equation}
or equivalently solving for $J$ as a function of $r_{\rm ph}$
\begin{equation}\label{Jofrph}
J=\frac{r_{\rm ph}\left(3M-r_{\rm ph}\right)}{6\sqrt{1-2M/r_{\rm ph}}} .
\end{equation}
From (\ref{rphradical}), the left side is negative on $r\in(2M,3M)$, zero at $r=3M$, and tends to $-2M^{2}$ as $r\to2M^{+}$. The right side is always negative and decreases monotonically as $J$ increases. Therefore the prograde photon sphere always lies in
\begin{equation}
2M<r_{\rm ph}<3M ,
\end{equation}
and decreases monotonically as $J$ increases, approaching $2M$.

A stable circular orbit requires the radial potential to have a maximum at the circular orbit. The innermost stable circular orbit, ISCO, is its critical point, equivalent to the circular orbit family (\ref{EL}) satisfying
\begin{equation}\label{isco}
\frac{dL}{dr}=0\qquad\left(\text{equivalent to}\ \frac{dE}{dr}=0\right) .
\end{equation}
Substituting (\ref{EL}) and simplifying, we obtain the following algebraic equation with a square root
\begin{equation}\label{iscoradical}
6J\left(r^{4}+2Mr^{3}+36J^{2}\right)\sqrt{Mr^{3}+9J^{2}}
=-Mr^{7}+6M^{2}r^{6}+18J^{2}r^{4}+72MJ^{2}r^{3}+648J^{4} .
\end{equation}
When $J\to0$, equation (\ref{iscoradical}) reduces to $Mr^{6}\left(6M-r\right)=0$, that is, the Schwarzschild result $r_{\rm ms}=6M$.

\begin{figure}[htbp]
  \centering
  \includegraphics[width=0.8\textwidth]{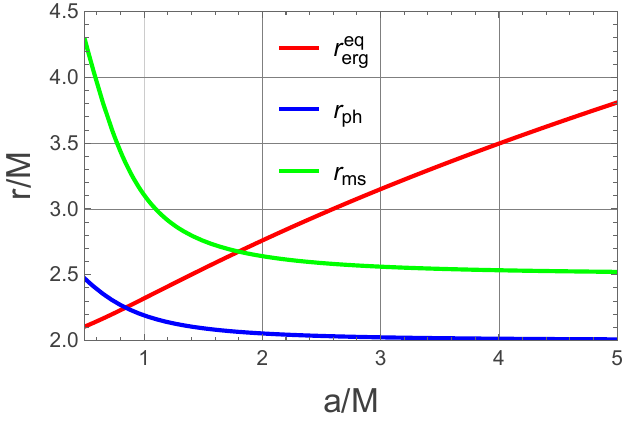}
  \caption{Characteristic radii of the Teo wormhole, prograde branch, as functions of spin $a/M$. The three curves are obtained by numerically solving equations (\ref{ergeq}), (\ref{rphradical}), and (\ref{iscoradical}). They are $r_{\rm erg}^{\rm eq}$, equatorial ergosphere boundary, $r_{\rm ph}$, prograde photon sphere, and $r_{\rm ms}$, innermost stable circular orbit.}
  \label{fig:2}
\end{figure}

Figure \ref{fig:2} shows the ergosphere boundary, photon sphere radius, and ISCO as functions of spin.
It can be seen from the figure that the prograde photon sphere rapidly shrinks to near $r=2M$ with rotation. $r_{\rm ph}$ intersects the ergosphere at $a/M=0.84375$, indicating that for $a/M<0.84375$ there is no circular orbit motion inside the ergosphere, and only for $a/M>0.84375$ can there be circular orbit motion. $r_{\rm ms}$ decreases monotonically with increasing spin to about $2.5M$, while $r_{\rm erg}^{\rm eq}$ increases monotonically. The two intersect at $a/M\simeq1.80$. When $a/M\lesssim1.80$, $r_{\rm ms}>r_{\rm erg}^{\rm eq}$, and all circular orbits inside the ergosphere are unstable. When $a/M\gtrsim1.80$, $r_{\rm ms}<r_{\rm erg}^{\rm eq}$, and stable circular orbits appear inside the ergosphere.

\section{The Comisso-Asenjo Magnetic Reconnection Process and Energy Extraction}

In this section we introduce the Comisso-Asenjo magnetic reconnection mechanism \cite{6}. The rotation of the wormhole drags magnetic field lines near the equatorial plane through $\omega=2J/r^{3}$, causing the radial component of the magnetic field to reverse sign on the two sides of the equatorial plane, thereby squeezing a thin current sheet between them. When the aspect ratio of the current sheet exceeds a critical value, it is disrupted by the plasmoid instability \cite{7,8}. Reconnection then proceeds rapidly at many $X$ points, converting magnetic energy into plasma kinetic energy and ejecting it. After reconnection ends, one branch corotating with the wormhole is accelerated, and the other branch is decelerated. Once the decelerated branch carries negative energy at infinity and passes through the throat into the other universe, it permanently disappears from our side of the spacetime. By energy conservation, the accelerated branch escapes to infinity carrying more energy than it originally had. Since reconnection is a local process, its dynamics is controlled by the dominant $X$ point, that is, the intersection of the separatrices that divide the global reconnection flow. Therefore the $r$ appearing below all refers to the radius $r_{X}$ of this $X$ point.

Intuitively, $\varepsilon_{\pm}$ is simplest to calculate in the frame that does not rotate with the wormhole. This frame is the zero angular momentum observer, ZAMO, frame \cite{9}. Transform from Boyer-Lindquist coordinates $(t,r,\theta,\varphi)$ to ZAMO coordinates $(\hat{t},\hat{x}^{1},\hat{x}^{2},\hat{x}^{3})$
\begin{equation}\label{zamoline}
ds^{2}=-d\hat{t}^{2}+\left(d\hat{x}^{1}\right)^{2}+\left(d\hat{x}^{2}\right)^{2}+\left(d\hat{x}^{3}\right)^{2},
\qquad
d\hat{t}=\kappa\,dt,\qquad d\hat{x}^{i}=\sqrt{g_{ii}}\,dx^{i}-\kappa\beta^{i}dt ,
\end{equation}
where $\kappa$ is the lapse function and $\beta^{i}$ is the shift. From $g^{\hat{t}\hat{t}}=-1$ and $g^{\hat{t}\hat{i}}=0$, we solve
\begin{equation}\label{zamo}
\kappa=\sqrt{-g_{tt}+\frac{g_{t\varphi}^{2}}{g_{\varphi\varphi}}},\qquad
\omega^{\varphi}=-\frac{g_{t\varphi}}{g_{\varphi\varphi}},\qquad
\beta^{\varphi}=\frac{\sqrt{g_{\varphi\varphi}}\,\omega^{\varphi}}{\kappa},\qquad
\beta^{r}=\beta^{\theta}=0 .
\end{equation}
Equation (\ref{zamo}) holds for any stationary axisymmetric metric. Substituting (\ref{ansatz}) gives two simplifications that belong only to this model. First, from $g_{t\varphi}=-r^{2}\omega\sin^{2}\theta$ and $g_{\varphi\varphi}=r^{2}\sin^{2}\theta$, we immediately get
\begin{equation}\label{wphi}
\omega^{\varphi}=\omega=\frac{2J}{r^{3}},
\end{equation}
that is, the frame dragging angular velocity is exactly the metric function $\omega$ itself. Second, substituting $g_{tt}=-N^{2}+r^{2}\omega^{2}\sin^{2}\theta$ into the first expression, the square term of $\omega$ cancels exactly
\begin{equation}\label{kappa}
\kappa=\sqrt{N^{2}-r^{2}\omega^{2}\sin^{2}\theta+r^{2}\omega^{2}\sin^{2}\theta}=N=\sqrt{1-\frac{2M}{r}} .
\end{equation}
Therefore the lapse function is independent of spin. The redshift factor is completely determined by $M/r$, the same as in Schwarzschild spacetime, and all effects of rotation are concentrated in the shift. From (\ref{kappa}) it is also seen that as long as $r>2M$, one has $\kappa^{2}>0$. Since (\ref{r0bound}) already guarantees $r\geq r_{0}>2M$, the ZAMO frame is defined in the entire exterior region of the wormhole. The shift on the equatorial plane is
\begin{equation}\label{beta}
\beta^{\varphi}=\frac{r\,\omega}{N}=\frac{2J}{r^{2}\sqrt{1-2M/r}} .
\end{equation}
The current sheet moves with the plasma at the Keplerian angular velocity $\Omega_{K}$ on a circular orbit. Its azimuthal velocity in the ZAMO frame is
\begin{equation}\label{vKdef}
\hat{v}_{K}=\frac{\sqrt{g_{\varphi\varphi}}}{\kappa}\,\Omega_{K}-\beta^{\varphi}
=\frac{r}{\kappa}\left(\Omega_{K}-\omega^{\varphi}\right) .
\end{equation}
Using (\ref{Omminusomega}) and (\ref{kappa}), the Keplerian velocity on the equatorial plane takes the very compact closed form
\begin{equation}\label{vK}
 \hat{v}_{K}=\frac{\mathcal{Q}}{r^{2}N},\quad
\mathcal{Q}=\sqrt{Mr^{3}+9J^{2}}-3J .
\end{equation}
Accordingly, the Lorentz factor of $\hat{v}_{K}$ is
\begin{equation}\label{gammaK}
\hat{\gamma}_{K}=\frac{1}{\sqrt{1-\hat{v}_{K}^{2}}}=\frac{r^{2}N}{\sqrt{\mathcal{D}}} ,
\end{equation}
where $\mathcal{D}$ is given by (\ref{Dpm}). From (\ref{gammaK}), $\hat{v}_{K}<1$ is completely equivalent to $\mathcal{D}>0$, that is, the range in which the Keplerian velocity in the ZAMO frame is meaningful is exactly the range in which circular orbits exist, (\ref{existr}).

We adopt the single-fluid approximation and assume that the plasma is an adiabatic, incompressible relativistic hot fluid. Since the efficiency of magnetic reconnection is extremely high, almost all magnetic energy in the reconnection region is converted into plasma kinetic energy. Therefore the electromagnetic part can be omitted in the energy-momentum tensor, and one can write
\begin{equation}\label{Tmunu}
T^{\mu\nu}=p\,g^{\mu\nu}+\omega_{0}U^{\mu}U^{\nu},
\end{equation}
where $p$, $\omega_{0}$, and $U^{\mu}$ are the proper pressure, enthalpy density, and four-velocity respectively. For a relativistic hot fluid, $\omega_{0}=4p$. The energy density at infinity is $e^{\infty}=-\kappa\,g_{\mu 0}T^{\mu 0}=\kappa\,\hat{e}+\kappa\beta^{\varphi}\hat{P}^{\varphi}$, where $\hat{e}$ and $\hat{P}^{\varphi}$ are the energy density and azimuthal momentum density in the ZAMO frame. Dividing it by the enthalpy density gives the energy density per unit enthalpy
\begin{equation}\label{epsdef}
\varepsilon\equiv\frac{e^{\infty}}{\omega_{0}} .
\end{equation}
Thus the energy densities per unit enthalpy of the accelerated and decelerated branches can be written as \cite{6}
\begin{equation}\label{epspm}
\varepsilon_{\pm}
=\kappa\hat{\gamma}_{K}\left[\left(1+\beta^{\varphi}\hat{v}_{K}\right)\sqrt{1+\sigma}
\ \pm\ \cos\xi\left(\beta^{\varphi}+\hat{v}_{K}\right)\sqrt{\sigma}
-\frac{1}{4\hat{\gamma}_{K}^{2}}\cdot
\frac{\sqrt{1+\sigma}\mp\cos\xi\,\hat{v}_{K}\sqrt{\sigma}}
{1+\sigma-\cos^{2}\xi\,\hat{v}_{K}^{2}\sigma}\right].
\end{equation}
Here $\sigma$ is the upstream magnetization parameter of the plasma, and $\xi$ is the angle between the magnetic field and the azimuthal direction in the local rest frame. The upper sign corresponds to the accelerated branch, and the lower sign corresponds to the decelerated branch. Therefore $\varepsilon_{\pm}$ is completely determined by the four quantities $\{a/M,r_{X}/M,\xi,\sigma\}$. Without loss of generality, in this paper we set $M=1$, so (\ref{epspm}) can be written as
\begin{equation}\label{epsexpanded}
\begin{aligned}
\varepsilon_{\pm} &= \frac{ \left[ r(r-2) + \frac{2J(\sqrt{r^3+9J^2}-3J)}{r^2} \right] \sqrt{1+\sigma} \pm \cos\xi \sqrt{1-\frac{2}{r}} (\sqrt{r^3+9J^2}-J) \sqrt{\sigma} }{ \sqrt{r^4 - 3r^3 - 18J^2 + 6J\sqrt{r^3+9J^2}} } \\
&\quad - \frac{ \sqrt{r^4 - 3r^3 - 18J^2 + 6J\sqrt{r^3+9J^2}} }{ 4r^2 } \frac{ \sqrt{1+\sigma} \mp \cos\xi \frac{\sqrt{r^3+9J^2}-3J}{r^2\sqrt{1-2/r}} \sqrt{\sigma} }{ 1 + \sigma - \cos^2\xi \frac{(\sqrt{r^3+9J^2}-3J)^2}{r^4(1-2/r)} \sigma }.
\end{aligned}
\end{equation}
By analogy with the Penrose process \cite{10}, only when the decelerated branch has negative energy at infinity can the accelerated branch obtain a net gain. For a relativistic hot fluid, the polytropic index is $\Gamma=4/3$ and the enthalpy density is $\omega_{0}=4p$, so energy extraction requires
\begin{equation}\label{cond}
\varepsilon_{-}<0,
\qquad
\Delta\varepsilon_{+}\equiv\varepsilon_{+}
-\left[1-\frac{\Gamma p}{(\Gamma-1)\omega_{0}}\right]=\varepsilon_{+}>0 .
\end{equation}
That is, the energy at infinity of the decelerated branch must be negative, and the energy at infinity of the accelerated branch must be positive. In practical calculations $\varepsilon_{+}$ is always positive, so the key criterion for determining whether energy extraction can occur is $\varepsilon_{-}<0$. In the Teo wormhole of this paper, the decelerated branch carrying negative energy passes through the throat $r=r_{0}$ into the other universe. It should be emphasized that formula (\ref{epspm}) has the same form as the Kerr black hole case, where $\varepsilon_{\pm}$ also depends only on $\{a/M,r_{X}/M,\xi,\sigma\}$. However, the functional form of the expansion (\ref{epsexpanded}) is completely different from the Kerr result. This is because in this paper $\kappa=N$ is independent of spin, and all rotation effects are concentrated in $\beta^{\varphi}$. The throat radius $r_{0}$ does not enter $\varepsilon_{\pm}$. It only acts indirectly through the domain $r_{0}\leq r_{X}<r_{\rm erg}^{\rm eq}$. Therefore the later analysis does not include the throat radius and by default takes the throat $r_0\to2M^+$. For a throat larger than 2, for example $r_0=3$, the plotted figures exclude the range $2<r<3$.

\begin{figure}[htbp]
  \centering
  \includegraphics[width=0.8\textwidth]{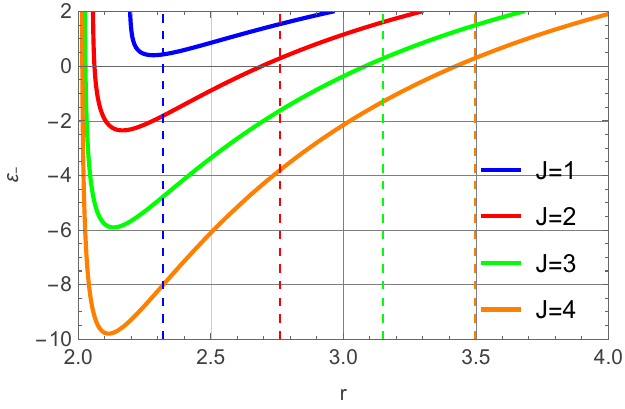}
  \caption{Solid lines represent $\varepsilon_{-}$ versus reconnection radius $r$ for different angular momenta. We take $\sigma=100$ and $\xi=\pi/12$. Dashed lines represent the ergosphere boundary for different angular momenta.}
  \label{fig:3}
\end{figure}

We plot $\varepsilon_{-}$ as a function of reconnection radius $r$ for different angular momenta in figure \ref{fig:3}.
It can be seen that when $J=1$, that is, when the wormhole rotates slowly, $\varepsilon_{-}$ is always greater than 0, and energy cannot be extracted. As angular momentum increases, the curve moves downward, and $\varepsilon_{-}$ becomes less than 0 at some reconnection radii, so energy extraction can proceed. The higher $J$ is, the more negative $\varepsilon_{-}$ becomes, which preliminarily indicates that higher angular momentum is more favorable for energy extraction. On the other hand, $\varepsilon_{-}$ at the ergosphere boundary is always greater than 0, showing that the conclusion that energy extraction occurs inside the ergosphere remains unchanged.

Next, we plot the allowed region for energy extraction in the $r-J$ plane, that is, the region with $\varepsilon_{-}<0$.
\begin{figure}[!h]
  \centering
  \begin{subfigure}{0.45\textwidth}
    \centering
    \includegraphics[width=\linewidth]{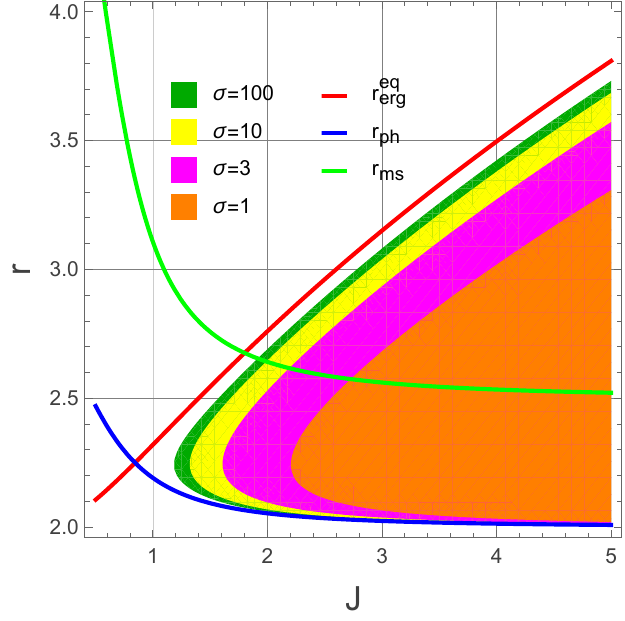}
  \end{subfigure}
  \begin{subfigure}{0.45\textwidth}
    \centering
    \includegraphics[width=\linewidth]{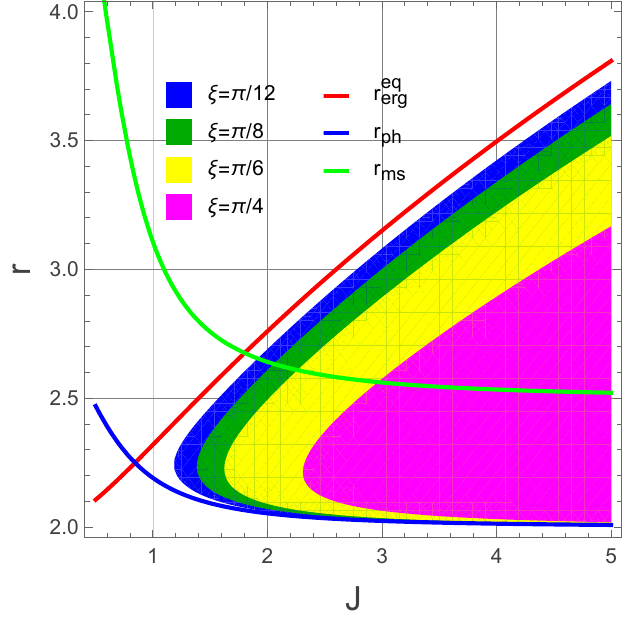}
  \end{subfigure}
\caption{Allowed region for energy extraction in the $r-J$ plane. Blue, red, green solid lines represent photon sphere, ergosphere boundary, ISCO. Left panel fixed $\xi=\pi/12$, different $\sigma$. Right panel fixed $\sigma=100$, different $\xi$.}
\label{fig:4}
\end{figure}
From figure \ref{fig:4}, the allowed region for energy extraction lies between the photon sphere and the ergosphere. Larger $\sigma$ and smaller $\xi$ make the allowed region larger, so energy extraction is more favorable. This is consistent with the conclusion of reference \cite{6}.

\begin{figure}[htbp]
  \centering
  \includegraphics[width=0.8\textwidth]{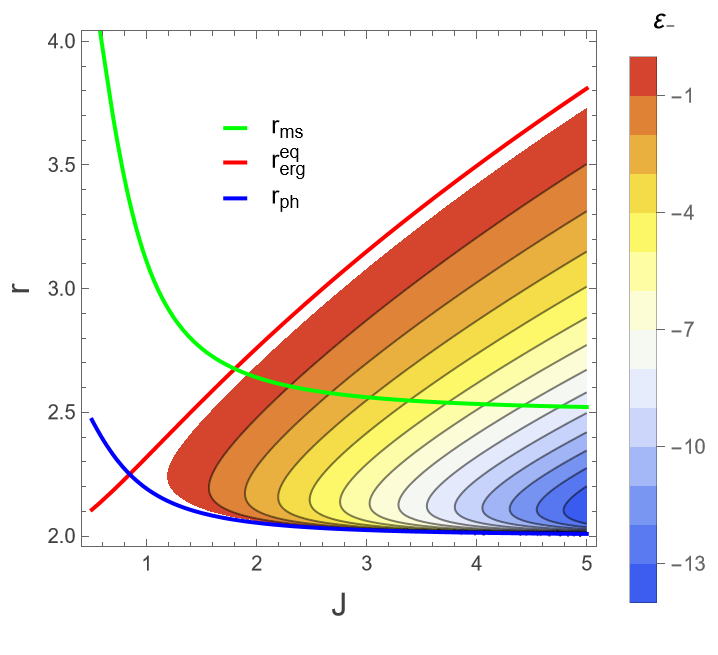}
  \caption{Distribution of $\varepsilon_{-}$ in the $(J,r)$ plane, with $\sigma=100$ and $\xi=\pi/12$.
    The color bar gives the value of $\varepsilon_{-}$, and the white region corresponds to $\varepsilon_{-}\geq0$, that is, no energy extraction.
    Blue, red, green solid lines represent photon sphere, ergosphere boundary, ISCO.}
  \label{fig:5}
\end{figure}

We plot $\varepsilon_{-}$ as a two-dimensional field in the $(J,r)$ plane in figure \ref{fig:5}, with color indicating the value of $\varepsilon_{-}$.
It can be seen that the absolute value of $\varepsilon_{-}$ increases with angular momentum, again showing that larger angular momentum is more favorable for energy extraction. The most suitable reconnection radius is about $r\approx2.1-2.2$, near the photon sphere.

\section{Power and Efficiency of Energy Extraction}
In this section we analyze the power and efficiency of energy extraction. The power per unit enthalpy of energy extraction is \cite{6}
\begin{equation}\label{Pextr}
P_{\rm extr}=-\varepsilon_{-}\omega_0A_{\rm in}\,U_{\rm in} ,
\end{equation}
where $A_{\rm in}$ is the incident cross-section of the swallowed plasma.
For the collisionless and collisional cases, $U_{\rm in}=O(10^{-1})$ \cite{11} and $O(10^{-2})$ \cite{12}, respectively.
In this paper we take the collisionless case $U_{\rm in}=0.1$. For the Teo wormhole of this paper, $A_{\rm in}$ can be expressed as
\begin{equation}\label{Ain}
A_{\rm in}\sim\left(r_{\rm erg}^{\rm eq}\right)^{2}-r_{\rm in}^{2} ,
\quad
r_{\rm in}=\begin{cases}
r_{\rm ph}, \ \ (\text{photon sphere exists outside the throat}) ,\\[2pt]
r_{0},      \ \ (\text{no photon sphere outside the throat}) .
\end{cases}
\end{equation}
Since this paper by default takes $r_{0}\to2M^{+}$, formula (\ref{Ain}) takes the first branch.

\begin{figure}[!h]
  \centering
  \begin{subfigure}{0.45\textwidth}
    \centering
    \includegraphics[width=\linewidth]{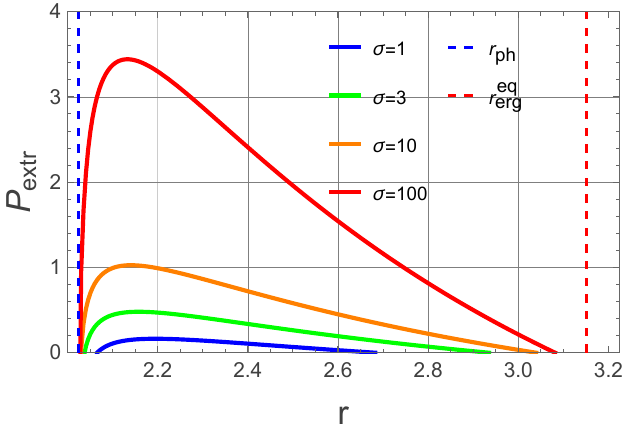}
  \end{subfigure}
  \begin{subfigure}{0.45\textwidth}
    \centering
    \includegraphics[width=\linewidth]{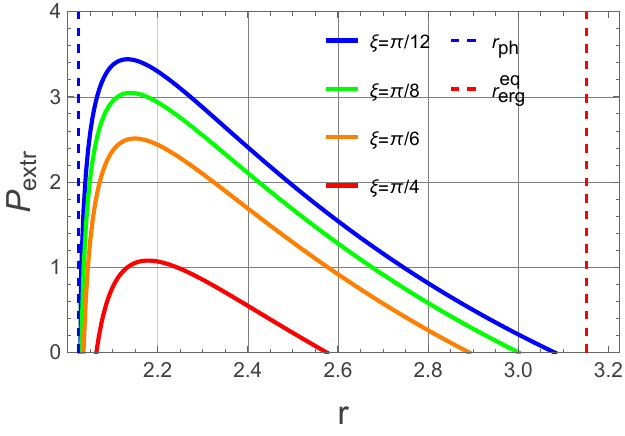}
  \end{subfigure}
\caption{Energy extraction power versus $r$. We take $J=3$. Blue dashed line represents photon sphere radius, red dashed line represents ergosphere boundary. Left panel fixed $\xi=\pi/12$, varying $\sigma$. Right panel fixed $\sigma=100$, varying $\xi$.}
\label{fig:6}
\end{figure}
We plot the energy extraction power as a function of $r$ in figure \ref{fig:6}. It can be seen that as $r$ increases, the power first increases and then decreases. Larger $\sigma$ or smaller $\xi$ gives larger energy extraction power, so energy extraction is more favorable. This is the same as the previous analysis of the allowed region and is also consistent with the Kerr black hole conclusion of reference \cite{6}.

\begin{figure}[htbp]
  \centering
  \includegraphics[width=0.8\textwidth]{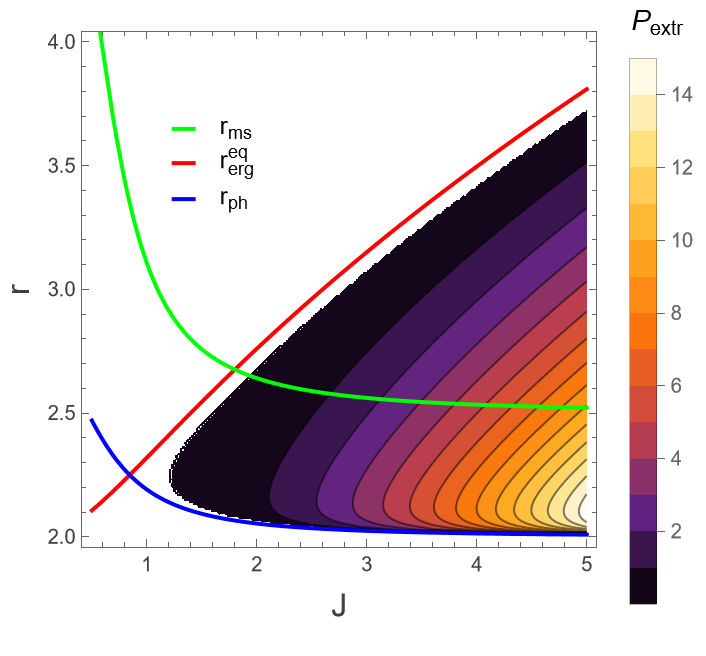}
  \caption{Power distribution for energy extraction. We take $\sigma=100$ and $\xi=\pi/12$.
    The color bar gives the power value. Blue, red, green solid lines represent photon sphere, ergosphere boundary, ISCO.}
  \label{fig:7}
\end{figure}
We plot the power distribution for energy extraction in the $(J,r)$ plane in figure \ref{fig:7}. Figure \ref{fig:7} again shows that the higher the angular momentum, the higher the energy extraction power, so magnetic reconnection energy extraction in the Teo wormhole is more favorable. This is completely different from the Simpson-Visser rotating wormhole in reference \cite{3}, where too large a spin causes the power to decrease.

On the other hand, the energy extraction power of the Teo wormhole is stronger than that of the Kerr black hole. Note that we cannot compare the Kerr black hole power point by point. For example, the reconnection radius $r$ for magnetic reconnection energy extraction in the Teo wormhole is above 2, while for the Kerr black hole $r$ is below 2, because the Kerr black hole ergosphere is below 2. The Teo wormhole spin can be arbitrarily high, while the Kerr black hole spin is from 0 to 1. We compare the power from a global perspective. As shown in figure \ref{fig:7}, the Teo wormhole power can reach 14, while under the same $\sigma$ and $\xi$, the extreme Kerr black hole power does not exceed 2, see figure 5 of reference \cite{6}. Figure 5 of reference \cite{6} takes spin $a=0.99$, and the power is about 0.71. Calculation gives an upper limit of about 1.6 for $a\to1$. This shows that the Teo wormhole has higher power than the Kerr black hole.

\begin{figure}[htbp]
  \centering
  \includegraphics[width=0.8\textwidth]{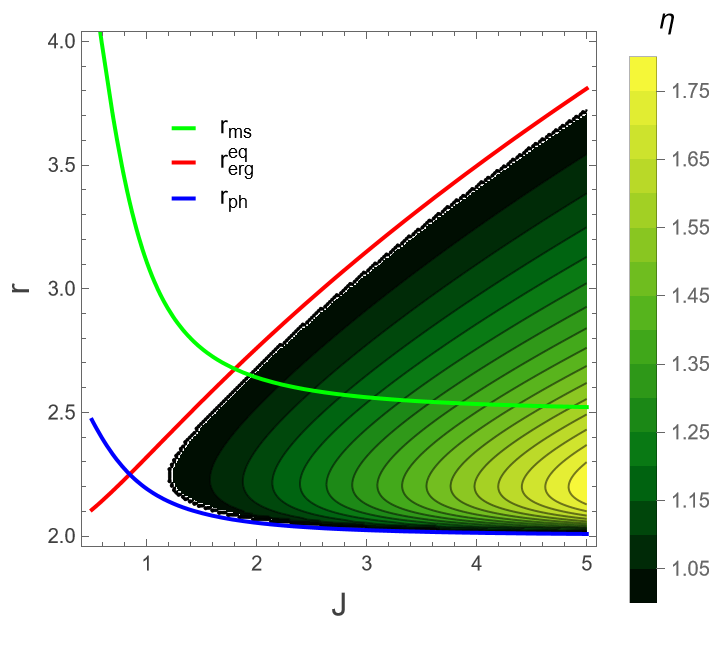}
  \caption{Efficiency distribution for energy extraction. We take $\sigma=100$ and $\xi=\pi/12$.
    The color bar gives the efficiency value. Blue, red, green solid lines represent photon sphere, ergosphere boundary, ISCO.}
  \label{fig:8}
\end{figure}

The energy extraction efficiency is \cite{6}
\begin{equation}
\eta=\frac{\varepsilon_+}{\varepsilon_++\varepsilon_-}.
\end{equation}
For energy extraction to be meaningful, the efficiency must be greater than 1. Similar to figure \ref{fig:7}, we plot the efficiency distribution for energy extraction in figure \ref{fig:8}. From the efficiency figure, the higher the angular momentum, the higher the efficiency. Therefore, according to the efficiency distribution figure, it can also be shown that larger angular momentum is more favorable for energy extraction. The most suitable reconnection radius is about $r\approx2.1-2.2$.
Of course, the efficiency also needs to be compared with the Kerr black hole. As shown in figure \ref{fig:8}, the efficiency of the Teo wormhole can reach about 1.75. According to figure 6 of reference \cite{6}, the maximum efficiency of the Kerr black hole is about 1.5. Their magnetization parameter is the same as in figure \ref{fig:8}, but their orientation angle is smaller than in figure \ref{fig:8}. If their orientation angle $\xi$ is also taken as $\pi/12$, then the maximum efficiency cannot reach 1.5. Calculation gives an upper limit of about 1.46 for $a\to1$. This shows that the efficiency of the Teo wormhole is stronger than that of the Kerr black hole.

\section{Energy Extraction in the Plunging Region}

In the previous sections, magnetic reconnection was discussed in the circular orbit region. However, in the Teo wormhole of this paper, doing so excludes a large region that is completely physically accessible.
From (\ref{existr}), Keplerian motion on the equatorial plane exists only for $r>r_{\rm ph}$,
while the throat is located at $r_{0}>2M$ and $r_{\rm ph}$ rapidly shrinks to $2M$ as spin increases.
Therefore for any $r_{0}>2M$, there is no circular orbit in the region $r_{0}\leq r\leq r_{\rm ph}$,
and the circular orbit description is simply not applicable there. More notably, when $a/M<0.84375$, $r_{\rm ph}>r_{\rm erg}^{\rm eq}$,
and the entire ergosphere lies inside the photon sphere. At this time there is not even one circular orbit inside the ergosphere,
and according to the criteria in sections 4 and 5, energy extraction is completely impossible. This naturally raises a question.
If the plasma does not move along circular orbits, does the above conclusion still hold?

To answer this question, this section turns to another configuration, in which the plasma initially moves along a circular orbit outside the innermost stable circular orbit,
reaches $r_{\rm ms}$, and then, because the orbit is no longer stable, acquires a radial velocity and begins to plunge inward \cite{52,53,74}.
The region
\begin{equation}
r_{0}\leq r<r_{\rm ms}
\end{equation}
is called the plunging region. The plunging negative energy plasma eventually passes through the throat into the other universe. Since the plasma in the plunging region has a radial velocity, the Keplerian velocity formula (\ref{vK}) containing only the azimuthal component in section 4 is no longer applicable, and its velocity in the ZAMO frame needs to be reestablished.

After the plunge begins, $E$ and $L$ remain conserved, and their values are given by the circular orbit at the starting point of the plunge, that is, the ISCO. Define
\begin{equation}\label{EILI}
E_{I}\equiv E\left(r_{\rm ms}\right),\quad L_{I}\equiv L\left(r_{\rm ms}\right) ,
\end{equation}
both obtained by evaluating (\ref{EL}) at $r=r_{\rm ms}$, where $r_{\rm ms}$ satisfies (\ref{iscoradical}).
In the ZAMO frame, the four-velocity can be decomposed as \cite{52,53,54}
\begin{equation}
\hat{U}^{\mu}=\hat{\gamma}_{s}\left(1,\hat{v}_{s}^{(r)},0,\hat{v}_{s}^{(\varphi)}\right)
=\left(\frac{E-\omega^{\varphi}L}{\kappa},\ \sqrt{g_{rr}}\,U^{r},\ 0,\ \frac{L}{\sqrt{g_{\varphi\varphi}}}\right) .
\end{equation}
This expression holds for any stationary axisymmetric metric. Substituting (\ref{ansatz}), $g_{rr}=\left(1-r_{0}/r\right)^{-1}$,
$g_{\varphi\varphi}=r^{2}$ on the equatorial plane, $\kappa=N$, and $\omega^{\varphi}=\omega=2J/r^{3}$.
The radial component $U^{r}$ is given by the radial equation (\ref{radialeq}). On the equatorial plane, taking $\mu^{2}=1$ and $\mathcal{K}=L^{2}$,
\begin{equation}\label{Ur}
\left(U^{r}\right)^{2}=\left(1-\frac{r_{0}}{r}\right)
\left[\frac{\left(E-\omega L\right)^{2}}{N^{2}}-1-\frac{L^{2}}{r^{2}}\right] ,
\end{equation}
where the negative sign indicates inward motion. Therefore the factor $(1-r_{0}/r)$ in $\sqrt{g_{rr}}\,U^{r}$ exactly cancels with $g_{rr}$, giving
\begin{equation}\label{vsr}
\sqrt{g_{rr}}\,U^{r}
=-\sqrt{\frac{\left(E_{I}-\omega L_{I}\right)^{2}}{N^{2}}-1-\frac{L_{I}^{2}}{r^{2}}} .
\end{equation}
This is a simplification that belongs only to this model. The radial velocity in the ZAMO frame is independent of the throat radius $r_{0}$,
and the location of the throat only restricts the domain through $r\geq r_{0}$. Thus we obtain the three components of the ZAMO velocity of the plunging plasma, all of which contain only metric functions and $\left(E_{I},L_{I}\right)$
\begin{equation}
\begin{aligned}
\hat{\gamma}_{s}
  &= \frac{E_{I}-\omega L_{I}}{N},\quad\hat{v}_{s}^{(\varphi)}
  = \frac{L_{I}}{r\,\hat{\gamma}_{s}},\\
\hat{v}_{s}^{(r)}
  &= -\frac{1}{\hat{\gamma}_{s}}
     \sqrt{\frac{\left(E_{I}-\omega L_{I}\right)^{2}}{N^{2}}
           -1-\frac{L_{I}^{2}}{r^{2}}},\quad
\hat{v}_{s}= \sqrt{\left(\hat{v}_{s}^{(r)}\right)^{2}
           +\left(\hat{v}_{s}^{(\varphi)}\right)^{2}}
   = \sqrt{1-\hat{\gamma}_{s}^{-2}} .
\end{aligned}
\end{equation}
Substituting the above into the normalization condition, one can directly verify
$\hat{\gamma}_{s}^{2}\left[1-\left(\hat{v}_{s}^{(r)}\right)^{2}-\left(\hat{v}_{s}^{(\varphi)}\right)^{2}\right]=1$,
that is, $\hat{\gamma}_{s}=\left(1-\hat{v}_{s}^{2}\right)^{-1/2}$ is indeed the Lorentz factor of $\hat{v}_{s}$. With the two components of $\hat{v}_{s}$, the energy density per unit enthalpy at infinity of the reconnection outflow is \cite{6,52,53,54}
\begin{align}
\varepsilon_{\pm}^{\rm pl}
=\kappa\hat{\gamma}_{s}\gamma_{\rm out}\Bigg[
&\left(1+\beta^{\varphi}\hat{v}_{s}^{(\varphi)}\right)
\pm v_{\rm out}\left(\hat{v}_{s}+\beta^{\varphi}\frac{\hat{v}_{s}^{(\varphi)}}{\hat{v}_{s}}\right)\cos\xi
\nonumber\\
&\mp v_{\rm out}\,\beta^{\varphi}\,\frac{\hat{v}_{s}^{(r)}}{\hat{\gamma}_{s}\hat{v}_{s}}\sin\xi
\Bigg]
-\frac{\kappa}{4\hat{\gamma}_{s}\gamma_{\rm out}\left(1\pm\hat{v}_{s}v_{\rm out}\cos\xi\right)} ,\label{epspmpl}
\end{align}
where the outflow velocity and its Lorentz factor are
\begin{equation}\label{vout}
v_{\rm out}=\sqrt{\frac{\sigma}{1+\sigma}},\quad \gamma_{\rm out}=\sqrt{1+\sigma} .
\end{equation}
Therefore, after expanding (\ref{epspmpl}), we obtain
\begin{equation}
\begin{aligned}
\varepsilon_{\pm}^{\mathrm{pl}}
&= E_I\sqrt{1+\sigma} \pm \sqrt{\sigma}\cos\xi\,
\frac{\left(E_I-\frac{2J}{r^3}L_I\right)E_I-\left(1-\frac{2}{r}\right)}
{\sqrt{\left(E_I-\frac{2J}{r^3}L_I\right)^2-\left(1-\frac{2}{r}\right)}} \\
&\quad \pm \frac{2J}{r^3}\sqrt{\sigma}\sin\xi\,
\frac{\sqrt{r^2\left[\left(E_I-\frac{2J}{r^3}L_I\right)^2-\left(1-\frac{2}{r}\right)\right]-\left(1-\frac{2}{r}\right)L_I^2}}
{\sqrt{\left(E_I-\frac{2J}{r^3}L_I\right)^2-\left(1-\frac{2}{r}\right)}} \\
&\quad -\frac{1-\frac{2}{r}}
{4\left[\left(E_I-\frac{2J}{r^3}L_I\right)\sqrt{1+\sigma}
\pm\sqrt{\sigma}\cos\xi\sqrt{\left(E_I-\frac{2J}{r^3}L_I\right)^2-\left(1-\frac{2}{r}\right)}\right]}.
\end{aligned}
\end{equation}
Compared with the circular orbit region (\ref{epspm}), (\ref{epspmpl}) has an extra term proportional to $\hat{v}_{s}^{(r)}\sin\xi$,
which is exactly the new effect brought by the radial velocity. Importantly, when $\hat{v}_{s}^{(r)}\to0$ and $\hat{v}_{s}\to\hat{v}_{s}^{(\varphi)}=\hat{v}_{K}$,
(\ref{epspmpl}) exactly returns to the circular orbit result (\ref{epspm}). The plunging region and the circular orbit region share the same criterion. Energy extraction requires
\begin{equation}
\varepsilon_{-}^{\rm pl}<0 ,
\end{equation}
and the corresponding efficiency and power per unit enthalpy are still
\begin{equation}
\eta^{\rm pl}=\frac{\varepsilon_{+}^{\rm pl}}{\varepsilon_{+}^{\rm pl}+\varepsilon_{-}^{\rm pl}},\quad
P_{\rm extr}^{\rm pl}=-\varepsilon_{-}^{\rm pl}\omega_0A_{\rm in}^{\rm pl}\,U_{\rm in} .
\end{equation}
Regarding the cross-section $A_{\rm in}^{\rm pl}$, one adjustment is needed. The plunging plasma falls all the way to the throat and passes through it,
and is no longer blocked by the photon sphere as in the circular orbit case. Therefore
\begin{equation}\label{Ainplung}
A_{\rm in}^{\rm pl}\sim\left(r_{\rm erg}^{\rm eq}\right)^{2}-r_{0}^{2} .
\end{equation}
In the fast rotation limit $r_{\rm ph}\to2M$ and $r_{0}\to2M^{+}$, (\ref{Ainplung}) tends to (\ref{Ain}) in the circular orbit region,
so at high spin the powers of the two configurations can be directly compared.
When $a/M<0.84375$, $r_{\rm ph}>r_{\rm erg}^{\rm eq}$ makes (\ref{Ain}) negative and invalid, and only (\ref{Ainplung}) is meaningful.

Similarly, we plot in the $r-J$ plane the allowed region for energy extraction when the plunging region is included.
\begin{figure}[!h]
  \centering
  \begin{subfigure}{0.32\textwidth}
    \centering
    \includegraphics[width=\linewidth]{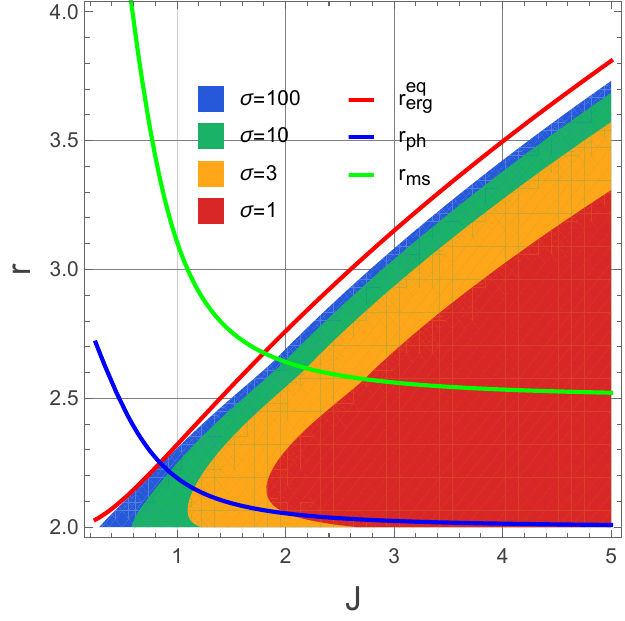}
  \end{subfigure}
  \begin{subfigure}{0.32\textwidth}
    \centering
    \includegraphics[width=\linewidth]{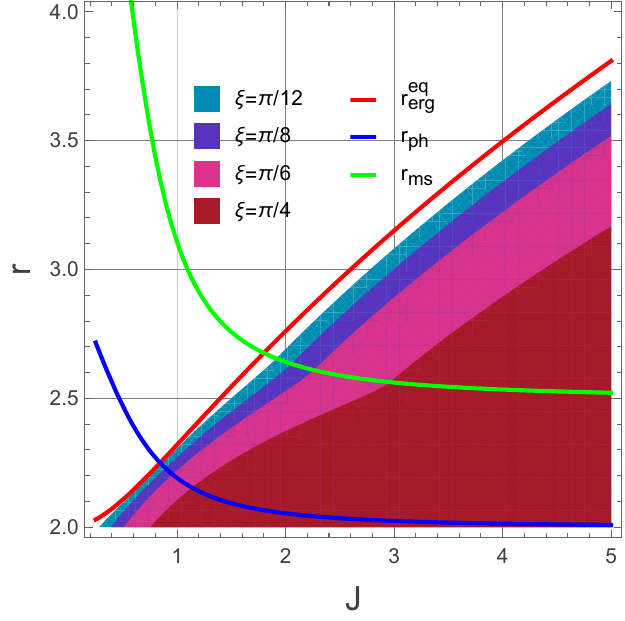}
  \end{subfigure}
  \begin{subfigure}{0.32\textwidth}
    \centering
    \includegraphics[width=\linewidth]{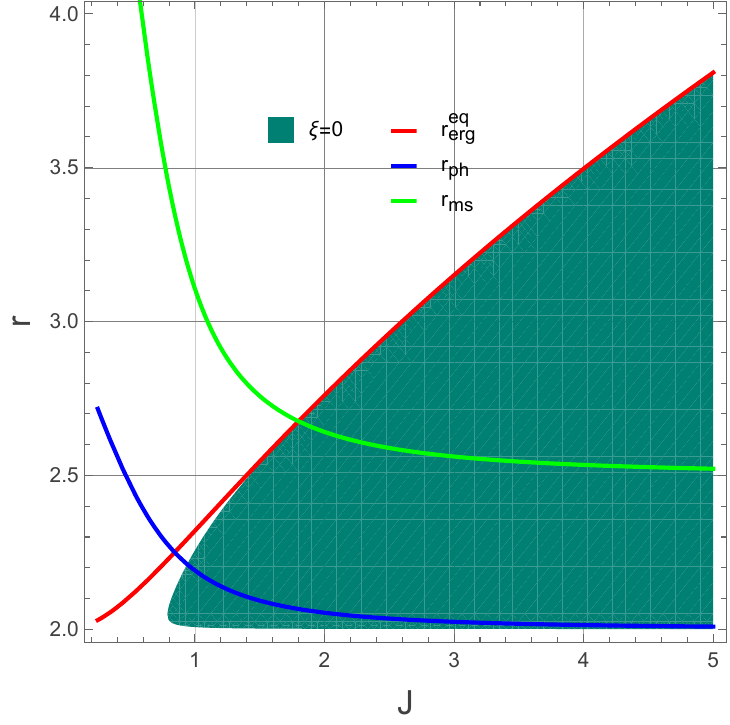}
  \end{subfigure}
\caption{Allowed region for energy extraction in the $r-J$ plane when the plunging region is included. Blue, red, green solid lines represent photon sphere, ergosphere boundary, ISCO. Left panel fixed $\xi=\pi/12$, different $\sigma$. Middle panel fixed $\sigma=100$, different $\xi$. Right panel fixed $\sigma=100,\xi=0$.}
\label{fig:9}
\end{figure}
From figure \ref{fig:9}, in the plunging region, larger $\sigma$ gives a larger allowed region for energy extraction. For $\xi$, when $\xi\neq0$, smaller $\xi$ gives a larger allowed region for energy extraction. When $\xi=0$, the allowed region for energy extraction changes. Compared with $\xi=\pi/12$, the allowed region for energy extraction decreases. This is a result specific to the orientation angle in the plunging region and is the same as in reference \cite{53}. In addition, compared with figure \ref{fig:4}, in the plunging region the allowed region for energy extraction is higher than in the circular orbit region.

\begin{figure}[!h]
  \centering
  \begin{subfigure}{0.32\textwidth}
    \centering
    \includegraphics[width=\linewidth]{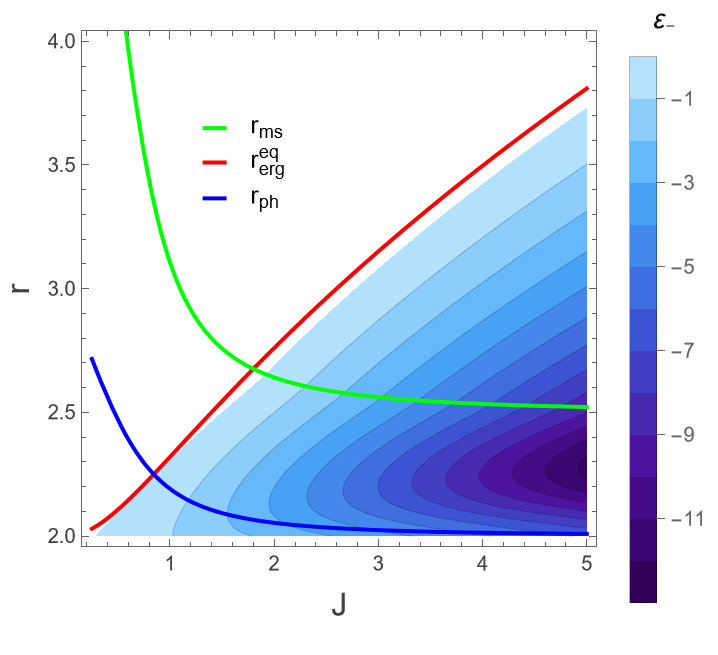}
  \end{subfigure}
  \begin{subfigure}{0.32\textwidth}
    \centering
    \includegraphics[width=\linewidth]{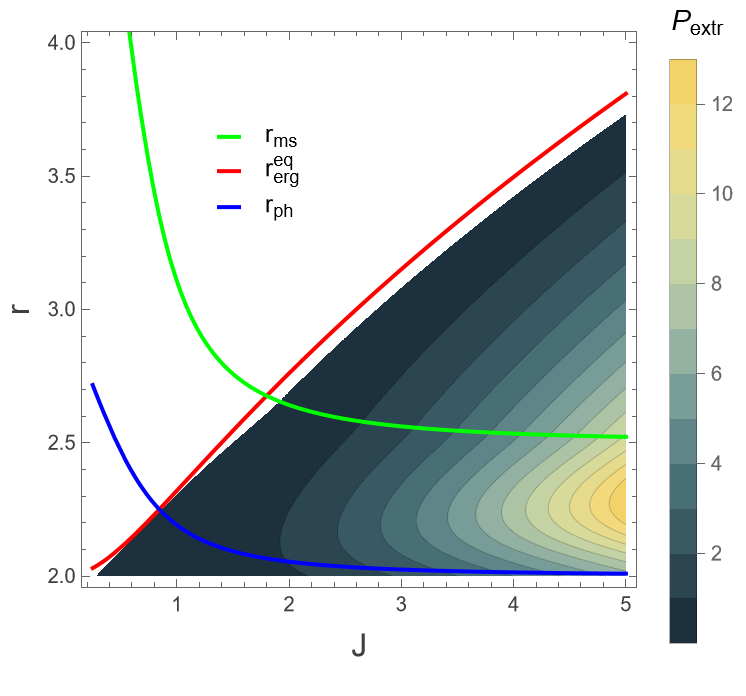}
  \end{subfigure}
  \begin{subfigure}{0.32\textwidth}
    \centering
    \includegraphics[width=\linewidth]{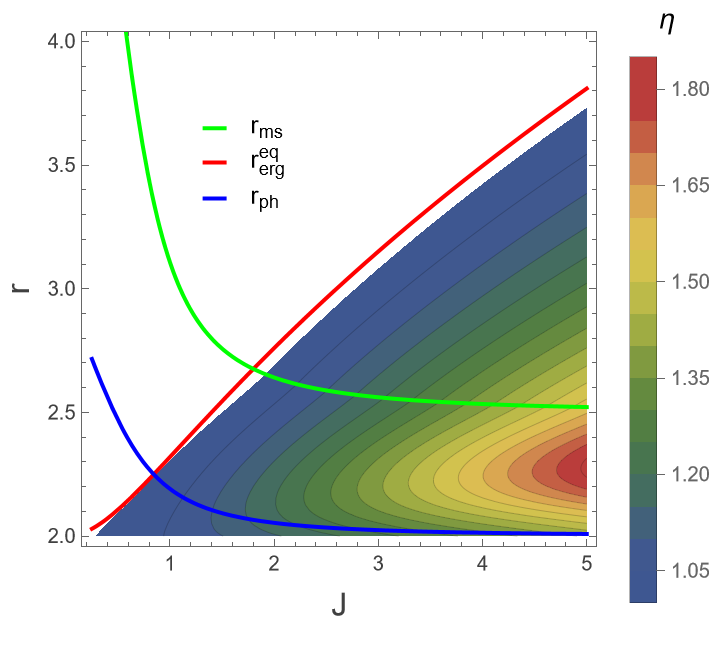}
  \end{subfigure}
\caption{Distributions in the $r-J$ plane of negative energy plasma energy density per unit enthalpy, power, and efficiency when the plunging region is included. Blue, red, green solid lines represent photon sphere, ergosphere boundary, ISCO. We fix $\sigma=100$ and $\xi=\pi/12$. When plotting the power distribution, to compare with the circular orbit region and maintain continuity, we adopt $A_{\rm in}\sim\left(r_{\rm erg}^{\rm eq}\right)^{2}-r_{0}^{2}$ in both the plunging region and the circular orbit region.}
\label{fig:10}
\end{figure}

Similarly, we plot in figure \ref{fig:10} the distributions in the $r-J$ plane of the negative energy plasma energy density per unit enthalpy, power, and efficiency when the plunging region is included.
It can be seen that in the plunging region, the most effective reconnection radius rises slightly to about $2.2-2.3$ as angular momentum increases. Compared with figures \ref{fig:5}, \ref{fig:7}, and \ref{fig:8}, within the plotted range $J\leq5$, the maximum values of the negative energy plasma energy density per unit enthalpy and the energy extraction power are lower than those in the circular orbit region, while the maximum value of the energy extraction efficiency is higher than that in the circular orbit region. This is clearly different from the Kerr black hole, where the power and efficiency in the plunging region are always higher than those in the circular orbit region \cite{53}.

To compare point by point the efficiency and power of the plunging region and the circular orbit region, we list table \ref{tab:2}.
\begin{table}[htbp]
\caption{Point by point comparison of energy extraction efficiency and power between the plunging region and the circular orbit region. We take $\sigma=100$, $\xi=\pi/12$, $r_{0}=2M$, and $U_{\rm in}=0.1$. The cross-sections of both configurations are uniformly taken as $A_{\rm in}\sim\left(r_{\rm erg}^{\rm eq}\right)^{2}-r_{0}^{2}$. $r_{X}^{\rm pl}$ and $r_{X}^{\rm circ}$ are the reconnection radii at which the efficiency reaches its maximum in that configuration.
$\Delta\eta\equiv\eta_{\max}^{\rm pl}-\eta_{\max}^{\rm circ}$, and $\Delta P\equiv P_{\max}^{\rm pl}-P_{\max}^{\rm circ}$.
N/A means that at that angular momentum the circular orbit region is empty, with $r_{\rm ph}>r_{\rm erg}^{\rm eq}$, or that although a circular orbit exists, $\varepsilon_{-}>0$ and energy extraction cannot occur. pl denotes plunging region, and circ denotes circular orbit region.}
\label{tab:2}
\begin{ruledtabular}
\begin{tabular}{cccccccccc}
$J/M^{2}$ & $r_{\rm ms}/M$ & $r_{X}^{\rm pl}/M$ & $\eta_{\max}^{\rm pl}$ & $P_{\max}^{\rm pl}$
          & $r_{X}^{\rm circ}/M$ & $\eta_{\max}^{\rm circ}$ & $P_{\max}^{\rm circ}$ & $\Delta\eta$ & $\Delta P$ \\
\hline
0.5 & 4.285 & 2.000 & 1.0157 & 0.013  & N/A   & N/A    & N/A    & N/A     & N/A     \\
1.0 & 3.107 & 2.006 & 1.0552 & 0.133  & N/A & N/A & N/A    & N/A     & N/A     \\
1.5 & 2.758 & 2.085 & 1.1152 & 0.468  & 2.231 & 1.0390 & 0.203  & $+0.076$ & $+0.265$ \\
2.0 & 2.641 & 2.136 & 1.1978 & 1.134  & 2.218 & 1.1232 & 0.851  & $+0.075$ & $+0.283$ \\
2.5 & 2.589 & 2.172 & 1.2916 & 2.164  & 2.211 & 1.2220 & 1.942  & $+0.070$ & $+0.222$ \\
3.0 & 2.561 & 2.201 & 1.3912 & 3.566  & 2.206 & 1.3288 & 3.499  & $+0.062$ & $+0.066$ \\
4.0 & 2.534 & 2.244 & 1.5990 & 7.492  & 2.201 & 1.5546 & 8.059  & $+0.044$ & $-0.567$ \\
5.0 & 2.522 & 2.276 & 1.8121 & 12.903 & 2.199 & 1.7889 & 14.576 & $+0.023$ & $-1.673$ \\
6.0 & 2.515 & 2.300 & 2.0276 & 19.784 & 2.197 & 2.0274 & 23.068 & $+0.000$ & $-3.283$ \\
8.0 & 2.509 & 2.334 & 2.4613 & 37.897 & 2.196 & 2.5110 & 45.999 & $-0.050$ & $-8.103$ \\
\end{tabular}
\end{ruledtabular}
\end{table}
From table \ref{tab:2}, the comparison results for efficiency and power are not the same.
First, for efficiency, the plunging region dominates when $J/M^{2}\lesssim6$, and $\Delta\eta>0$ decreases monotonically as angular momentum increases,
from $0.076$ at $J/M^{2}=1.5$ to $0.023$ at $J/M^{2}=5$. It changes sign at $J/M^{2}\simeq6.01$,
and at $J/M^{2}=8$, $\Delta\eta\simeq-0.050$, so the efficiency of the circular orbit region overtakes it.
Second, for power, the reversal comes earlier. $\Delta P$ changes sign at $J/M^{2}\simeq3.15$.
In the intermediate and low spin range $J/M^{2}\lesssim3.15$, $\Delta P>0$, and the plunging region power is instead higher.
The maximum gain appears at $J/M^{2}\simeq2$, with $\Delta P\simeq0.28$.
After that $\Delta P$ becomes negative and its absolute value increases rapidly. At $J/M^{2}=8$, the circular orbit region power is higher by about $8.1$.
Therefore the value of the plunging region is concentrated in the case where the spin is not particularly high.

Finally, we make a global comparison between the plunging region of the Teo wormhole and the plunging region of the Kerr black hole. We take $\sigma=100$ and $\xi=\pi/12$.
The cross-section of the Kerr black hole is taken as $A_{\rm in}\sim r_{\rm erg}^{2}-r_{+}^{2}$, and that of the Teo wormhole is taken as $A_{\rm in}\sim\left(r_{\rm erg}^{\rm eq}\right)^{2}-r_{0}^{2}$, with $r_{0}=2M$. In the near-extreme limit $a/M\to1$ of the Kerr black hole, $\eta_{\max}^{\rm pl}\simeq1.495$ and $P_{\max}^{\rm pl}\simeq1.72$, while in the same limit the circular orbit region gives $1.461$ and $1.60$.
The Teo wormhole has no such upper limit. Its spin can be arbitrarily large.
At $J/M^{2}=5$, the plunging region already reaches $\eta_{\max}^{\rm pl}\simeq1.812$ and $P_{\max}^{\rm pl}\simeq12.9$,
and at $J/M^{2}=8$ these further rise to $2.461$ and $37.9$. These results show that even compared with the limiting values of the Kerr black hole, the power and efficiency of the Teo wormhole plunging region are higher than those of the Kerr black hole.

\section{Conclusions}

This paper extends the Comisso-Asenjo magnetic reconnection mechanism from Kerr black holes and rotating wormholes that can reduce to Kerr to
the Teo traversable rotating wormhole, which has
no event horizon and cannot reduce to a Kerr black hole in any parameter limit.
  Taking the simplest regular form of this model that also carries ADM mass, (\ref{ansatz}), as an example, we systematically calculate the circular orbit structure on the equatorial plane, the criterion for energy extraction, the power, and the efficiency, and further extend the reconnection region from the circular orbit region to the plunging region.  

This model has at most one ergosurface, and there is no inner ergosurface like that in Kerr spacetime. The spin \(a/M\) has no upper bound like the Kerr black hole bound \(|a|\le M\). The photon sphere and ISCO radii decrease with increasing spin, while the equatorial ergosphere boundary increases with increasing spin. Numerical results show that the prograde photon sphere and the ergosphere boundary intersect at \(a/M=0.84375\). Below this value, there is no circular orbit inside the ergosphere.

After analyzing the magnetic reconnection process of this wormhole, we find that in both the circular orbit region and the plunging region, the negative energy plasma quantity $\varepsilon_-$ that determines whether energy extraction occurs is independent of the wormhole throat. It only acts indirectly through the domain. Because the domain lies between the wormhole throat and the ergosphere. Therefore all our analysis chooses the wormhole throat at $r_0\to2M^+$. In the circular orbit region, we find that larger magnetization and smaller orientation angle make energy extraction more favorable. This is the same as previous magnetic reconnection conclusions. In the Teo wormhole, we find that the most effective reconnection radius is stable at $r_{X}\approx2.1M$ to $2.2M$, almost independent of $J$, and both the energy extraction efficiency and power increase monotonically with angular momentum. By comparing with the Kerr black hole, the energy extraction power and efficiency of the Teo wormhole are higher than those of the Kerr black hole. It should be noted that this comparison can only be global and not point by point, because the parameter spaces of the two do not overlap.
In the plunging region, except for the special case $\xi=0$, larger magnetization or smaller orientation angle gives a larger allowed region for energy extraction, consistent with previous conclusions. We also find that the most effective reconnection radius increases monotonically with angular momentum, and both the energy extraction efficiency and power also increase monotonically with angular momentum. By global comparison with the Kerr black hole, the energy extraction power and efficiency of the Teo wormhole are higher than those of the Kerr black hole. By comparison with the circular orbit region, the plunging region is not comprehensively superior.

The Simpson-Visser type rotating wormhole studied in reference \cite{3} retains the three Kerr structure functions $\Sigma$, $\Delta$, and $A$,
and exactly reduces to a Kerr black hole when the regularization parameter $\ell\to0$.
The Teo wormhole in this paper is fundamentally different, producing three testable differences.
(i) The spin dependence is opposite. In reference \cite{3}, the large spin makes efficiency and power decrease,
and the efficiency exhibits a nonmonotonic peak with $\ell$. Here both efficiency and power increase monotonically with angular momentum, and there is no effect in which the large  spin is unfavorable.
(ii) There is no Kerr limit. Reference \cite{3} can return to the Kerr black hole at $\ell\to0$ and be compared with it point by point.
This model by construction contains no horizon, cannot reduce to Kerr, and can only be compared globally.
(iii) The inner boundary of the ergosphere is different. The inner boundary of the Kerr black hole ergosphere is the outer horizon $r_{+}$.
The fast rotating wormhole in reference \cite{3} is enclosed by inner and outer ergosurfaces together with the throat,
while the inner boundary of the ergosphere in this paper is the throat $r_{0}$, with neither an inner ergosurface nor a horizon.

This study of energy extraction from traversable rotating wormholes not only helps to understand the mechanism, efficiency, and power of energy extraction in wormhole spacetime, but also helps to understand the mechanism of energy extraction in horizonless compact objects. It is of important significance for a comprehensive understanding of the Comisso-Asenjo magnetic reconnection process.

This work extends the Comisso–Asenjo magnetic reconnection mechanism, investigating Kerr like black holes and rotating wormholes that can recover the Kerr geometry in an appropriate limit, to the Teo traversable rotating wormhole. The results reveal clear distinctions between wormhole and black hole geometries, providing a basis for further investigations into their energy-extraction mechanisms and potential intricate properties.

\noindent {\bf Acknowledgments}

\noindent
This work is supported by the National Natural Science Foundation of China (Grants Nos. 12375043, 12575069), Chongqing Normal University Fund Project (Grant No. 26XLB001), and Science and Technology Research Program of Chongqing Municipal Education Commission (Grant No. KJZD-K202600503).

\end{document}